\documentstyle [12pt,aasms]{article}
\input{epsf.tex}

\begin{document}

\title{The Velocity Dispersion -- Temperature Correlation from a Limited
Cluster Sample}
\author{Christina M. Bird}
\affil{Department of Physics and Astronomy, University of Kansas, Lawrence,
KS 66045}
\affil{tbird@kula.phsx.ukans.edu}
\author{Richard F. Mushotzky}
\affil{NASA/Goddard Space Flight Center,
Laboratory for High Energy Astrophysics,\\
Code 666, Greenbelt, MD 20771}
\affil{richard@xray-5.gsfc.nasa.gov}
\and
\author{Christopher A. Metzler}
\affil{Department of Physics, University of Michigan, Ann Arbor, MI
48109}
\affil{metzler@pablo.physics.lsa.umich.edu}
\vskip 4cm
\noindent{Submitted to the {\it Astrophysical Journal}}

\begin{abstract}

Most studies of correlations between X-ray and optical properties of
galaxy clusters have used the largest samples of data available,
regardless of the morphological types of clusters included.  Given the
increasing evidence that morphology is related to a cluster's degree of
dynamical evolution, we approach the study of X-ray and optical
correlations differently.  We evaluate the relationship between velocity
dispersion and temperature for a limited set of galaxy clusters taken
from Bird (1994), which all possess dominant central galaxies and which
have been explicitly corrected for the presence of substructure.  We
find that $\sigma _r \propto T^{0.61 \pm 0.13}$.   We use a Monte Carlo
computer routine to estimate the significance of this deviation from
the $\sigma _r \propto T^{0.5}$ relationship predicted by the virial
theorem.  We find that the
simulated correlation is steeper than the observed value only 4\% of the
time, suggesting that the deviation is significant.
The combination of protogalactic winds and dynamical friction
reproduces nearly exactly the
observed relationship between $\sigma _r$ and $T$.

\end{abstract}

\section{Introduction}

Galaxy clusters occupy a unique position in the dynamical evolution of the
universe.  Unlike lower-mass systems such as galaxies, which for the
most part retain little dynamical information about their formation,
clusters of galaxies are within one or two crossing times of their formation.
This suggests that they may retain valuable clues to their initial conditions
(as well as hints about the collapse and formation of structure in the
early universe).  The effect of the dense cluster environment on galaxy
evolution, as well as other
trends in the physical properties of clusters (see, for instance, Dressler
1984;
Giovanelli \& Haynes 1985; Edge \& Stewart 1991), suggests that they are
gravitationally bound and that their galaxies no longer participate in the
Hubble flow.  This distinguishes clusters from superclusters and other
large-scale structures.  The study of galaxy clusters thus provides a
unique opportunity to explore gravitational interactions and dynamical
evolution in the universe.

Clusters of galaxies contain two luminous components, hot gas and
galaxies.  If a cluster is sufficiently old and unperturbed, these tracer
particles will have equilibrated within the cluster gravitational potential.
This enables use of the equations of hydrostatic and dynamical equilibrium
to explore the physical properties of these systems.  For a hot gas in
equilibrium with a spherical gravitational potential, the equation of
hydrostatic equilibrium may be written
\begin{equation}
M_X(r) = { {-kTr} \over {G \bar{m}} } ({ {d~{\rm ln}~n_{gas}
} \over {d~{\rm ln}~r}
} + { {d~{\rm ln}~T} \over {d~{\rm ln}~r}} )
\end{equation}
(e.g.\ Fabricant, Lecar \& Gorenstein 1981), where $M_X$ is the X-ray
determined
virial mass, $T$ is the temperature of the X-ray emitting gas, $n_{gas}$
is the gas density and $\bar{m}$ is the average mass per gas particle.
Similarly, the Jeans
equation relates the kinetic energy of the galaxies to the virial mass of
the cluster:
\begin{eqnarray}
- {{G~n_{gal} M_{opt}(r)} \over r^2} & = & { d(n_{gal} \sigma ^2 _r) \over dr}
+
{2 n_{gal} \over { r \sigma _r ^2} } (1 - \sigma _r ^2 / \sigma ^2 _t) \\
\nonumber
 & = & { d(n_{gal} \sigma ^2 _r) \over dr} + {2 n_{gal} \over
{ r \sigma _r ^2} } A
\end{eqnarray}
(Merritt 1987), where $M_{opt}$ is the optically-determined virial mass,
$r$ is the clustercentric radius,
$n_{gal}$ is the galaxy density, $\sigma _r$ and $\sigma _t$ are the
radial and tangential velocity dispersions respectively, and $A$ is the
anisotropy parameter describing the distribution of galaxy orbits.

For an isothermal
cluster in dynamical equilibrium, with no source of energy other
than gravity,
the masses as determined by the galaxies and by the
gas are expected to be equal.  As shown by Bahcall \& Lubin (1994) among
others,
the ratio of the
kinetic energies of the galaxies and gas is then equal to the ratio of the
logarithmic slope of the gas density profile to that of the galaxies:
\begin{eqnarray}
{{\sigma ^2 _r} \over {kT \over \bar{m} } } & = &
{ {d~{\rm ln}~n_{gas} } / {d~{\rm ln}~r} \over
{d~{\rm ln}~n_{gal} } / {d~{\rm ln}~r + 2A} }. \\ \nonumber
& = & { {d~{\rm ln}~n_{gas} } / {d~{\rm ln}~r} \over
{d~{\rm ln}~n_{gal} } / {d~{\rm ln}~r}}
\end{eqnarray}
(where $A = 0$ for an isotropic distribution of galaxy orbits).
Therefore, using the assumptions that the gas and galaxies are both in
equilibrium with the cluster gravitational potential,
and that gravity is the only source of energy, allows us to predict
that the velocity dispersion (as measured from galaxy velocities) and the
temperature of the intracluster medium (as determined from X-ray spectra)
should be correlated, with $\sigma _r \propto T ^{0.5}$.  The ratio of the
kinetic energies is called $\beta _{spec}$.  The ratio of the
logarithmic slopes of the
density profiles is $\beta _{fit}$.

Despite the many difficulties in accurately measuring cluster temperatures
and velocity dispersions,
studies of X-ray and optical cluster samples reveal a well-behaved
correlation between these quantities (Mushotzky 1984; Edge \& Stewart 1991,
hereafter ES91;
Lubin \& Bahcall 1993, hereafter
LB93).  The relationship between $\sigma _r$ and $T$
expected from virial considerations is consistent with the data, although
there is a large scatter about the $\sigma _r \propto T^{0.5}$ line.
This scatter has been attributed to incomplete gas
thermalization, cooling flows, velocity anisotropies in the galaxy orbits,
foreground/background contamination, and substructure in the clusters
(cf.\ ES91; LB93 and references therein).

It is important to remember, however, that the predicted
$\sigma _r - T$ correlation derives from the virial theorem, and
that in order to test it one must consider the dynamical state of the clusters
in the dataset (cf.\ Gerbal et al.\ 1994).
The high frequency of substructure in clusters of all
morphologies, as determined by both X-ray and optical studies (see, e.g.\,
Davis \& Mushotzky 1993; Mohr, Fabricant \& Geller 1993; Beers et al.\ 1991;
Bird 1993, 1994), is generally believed to indicate that clusters are
dynamically-young.  If clusters are only within a few crossing times of
formation, then in many cases virial equilibrium has not been established.
This certainly influences the broad distribution of clusters about the
canonical
$\sigma _r \propto T^{0.5}$ relation.

In this paper we will quantify the effects of
morphology and substructure on the velocity
dispersion-temperature correlation for clusters.  In Section 2 we
present the limited cluster sample, in which the morphological type of the
cluster sample has been restricted and the effects of substructure
have been minimized.  We have supplemented the available published X-ray
temperature data with new, more
accurate temperatures from ASCA and {\it Ginga}.  In
Section 3 we present the regressions between the velocity dispersion and
temperature.  Section 4 summarizes proposed mechanisms for modifying the
slope of the $\sigma _r - T$ correlation.  In Section 5 we present a summary.

\newpage

\section{The Limited Cluster Sample}

The morphology of a cluster may be described by its gas and/or galaxy
distribution.  As our observations of clusters have improved,
it has become clear
that morphology is related to the dynamical age of a cluster.
Irregular clusters are dynamically young, and tend to be spiral-rich and
gas-poor.  They tend to have non-Gaussian velocity distributions and
kinematically-distinct subconcentrations of galaxies.
Regular clusters are dominated by ellipticals, have Gaussian velocity
distributions and tend
to be luminous X-ray emitters (cf.\ Sarazin 1988 and references therein;
Bird 1993,1994).

Bird (1994) presents a detailed analysis of the dynamics of
nearby clusters ($z < 0.1$) with
central galaxies.  These clusters tend to have smooth morphologies and
X-ray cooling flows, and in the past it has been assumed that they represent
the most relaxed, dynamically-evolved clusters in the universe.  However,
Bird (1994) shows that these clusters also possess significant
substructure.  An objective partitioning algorithm called KMM
(McLachlan \& Basford 1988; Ashman, Bird \& Zepf 1994) is used to
remove galaxies belonging to subsystems in the clusters, and the dynamical
properties of the ``cleaned'' (i.e., substructure corrected)
cluster datasets are presented.  It is the
25 clusters in this ``cD database'' which form the optical
sample of the present analysis.

Of the 25 clusters used in Bird (1994), 21 have
accurate X-ray temperature measurements.
These clusters, which will be referred to as the
limited cluster sample, are listed in Table 1. Table 1 includes the
following information:  column (1), the cluster name; (2), the 1-D velocity
dispersion of the cluster (estimated using the robust biweight estimator
$S_{BI}$, Beers, Flynn \& Gebhardt 1991) without substructure correction;
(3), the velocity dispersion corrected for substructure; (4), the X-ray
temperature, (5) the source code for the X-ray measurement.  The optical
redshifts are taken from the literature, with sources given in Bird (1994).
In addition we have added the Centaurus Cluster (A3526), which was
excluded from the cD study because of its proximity.  The X-ray temperatures
are taken from single-temperature models to ASCA or {\it Ginga} spectra
where available, and then
from EXOSAT and the {\it Einstein} MPC.  For the clusters A1736 and A3558,
the GINGA observations are best-fit by a two-temperature model (Day et al.\
1991), in contradiction to both the {\it Einstein} and ROSAT spectra.
Because the data are inconclusive, we have included both temperatures in
Table 1 for these two clusters, and we will consider them both in the
statistical analysis.

Note that
the velocity dispersion presented here is measured only along our line of
sight to the cluster.  We assume for the
moment that any velocity anisotropy in these
clusters is small and therefore $\sigma _{LOS}$ is comparable to $\sigma _r$
(we will explore this assumption in more detail below).

In Table 2 we present the
individual values of $\beta _{spec}$ for the limited cluster
sample, both with and without substructure correction.  With no substructure
correction, the mean value of $\beta$ is 1.20$^{+0.30}_{-0.18}$,
with an rms scatter of 0.66 (GINGA:  0.99$^{+0.24}_{-0.17}$, rms 0.43).
The high mean value and large scatter are due to the inclusion of A2052 in
the dataset.  The uncorrected velocity dispersion of this cluster is extremely
high, 1404 km s$^{-1}$, with corresponding $\beta _{spec} = 3.51$.
If this datapoint
is excluded from the list, the mean drops to 1.09$^{+0.15}_{-0.15}$
with rms scatter 0.43 (GINGA:  0.97$^{+0.24}_{-0.17}$, rms 0.42).
Including the substructure correction
to the velocity dispersion (and retaining A2052, which is no
longer anomolous), $\langle \beta _{spec} \rangle = 0.90^{+0.10}_{-0.15}$
with an rms scatter of 0.37 (where the confidence intervals are the 90\%
bootstrapped estimates) (GINGA:  0.87$^{+0.12}_{-0.17}$, rms 0.38) .

To demonstrate the effect of morphology on $\beta _{spec}$,
these numbers should be
compared to the values from the LB93 study.  Lubin \&
Bahcall use 41 clusters of widely varying morphology.  Their mean value of
$\beta _{spec}$ is 1.14$^{+0.08}_{-0.08}$
with an rms scatter of 0.57.  The ES91
sample, being based on an X-ray flux-limited catalog of clusters, is biased
toward X-ray luminous systems, which are less likely to be affected by major
substructure.  This sample yields $\langle \beta _{spec}
\rangle = 0.91^{+0.11}_{-0.13}$ with an rms scatter of 0.38.
It is clear that when examining correlations between temperature and
velocity dispersion, uncertainty may be introduced by
neglecting the effects of morphology and substructure in the dataset.

\section{The Velocity Dispersion -- Temperature Correlation}

In Figure 1, we present the velocity dispersion and temperature data for the
22 clusters in the limited sample.  The velocity dispersions are corrected
for substructure.  The dashed lines are the correlations predicted by the
virial theorem, for $\beta_{spec} = 1$ and for $\beta_{spec} = 0.67$.
Recall that for these data $\langle \beta _{spec}
\rangle
= 0.90$.  The solid line is the best fit to the data using the lower
temperatures for A1736 and A3558:
\begin{equation}
\sigma _r = 10^{2.50 \pm 0.09} T^{0.61 \pm 0.13}
\end{equation}
Similarly, we find that
\begin{equation}
T = 10^{-3.15 \pm 0.60} \sigma _r ^{1.31 \pm 0.21}
\end{equation}
For the higher GINGA temperatures for these two clusters, we find that
\begin{equation}
\sigma _r = 10^{2.39 \pm 0.09} T^{0.76 \pm 0.11}
\end{equation}
and
\begin{equation}
T = 10^{-3.21 \pm 0.61} \sigma _r ^{1.34 \pm 0.21}
\end{equation}
In both equations the uncertainties quoted are the bootstrapped 1-$\sigma$
values.
This fit includes the errors in the measurements, using a linear fitting
technique developed by Akritas, Bershady \& Bird (1995, in preparation).
This algorithm, based on the ordinary least-squares bisector first defined
by Isobe et al.\ (1990),
explicitly includes both intrinsic scatter in the relation and uncorrelated
measurement errors.   The bisector method assumes that neither variable is
dependent on the other, which is probably appropriate for the current
physical situation.  The velocity dispersion and X-ray temperature are both
determined by the depth of the gravitational potential (and perhaps other
physical effects), and are therefore {\it independent} of each other.

This subtlety in the application of linear regression algorithms has been
previously noted by astrophysicists for other applications, such as
the Tully-Fisher effect (see Isobe et al.\ 1990 for a detailed discussion),
but not yet applied to the problem of X-ray and optical correlations.
The use of an inappropriate or biased
regression technique can have a significant
effect on the coefficients of the linear fit, as we demonstrate in Table
3.  To simplify this discussion, in Table 3 we present the following:
\begin{itemize}
\item
the published linear regressions given in ES91 and LB93
\item
the linear regressions determined from an ordinary
least squares fit, without measurement errors
\item
the linear regressions from the bisector lines, with and
without measurement errors
\end{itemize}
for the ES91 and LB93 datasets, as well as similar regressions for our
limited cluster dataset.  The uncertainties in the linear coefficients
are the 1-$\sigma$ values, determined using a bootstrap method which is the
preferred estimator for small datasets.

First of all, we see that the published linear regressions are recovered
for both the ES91 and the LB93 datasets using the ordinary least squares (OLS)
regressions, without errors.  For these fits, the velocity dispersion is
assumed to be {\it dependent} on the temperature, which as discussed
above does not seem like a physically well-motivated assumption.
In addition, simulations suggest that the OLS regressions are severely
biased for such small sample sizes.
The bisector slopes for all three datasets are much steeper than the
OLS slopes, varying from 0.61 for our limited cluster dataset
and the {\it Einstein} data to 0.87 for
the LB93 dataset.  The regression for our limited cluster dataset is
marginally consistent with the slope of 0.5 predicted by the virial theorem.
For the ES91 and LB93 datasets, the fitted slopes are at least 3$\sigma$
away from the canonical value of 0.5.

Given the large dispersions between the individual linear regressions,
as well as
the coefficients of the regressions for the three datasets, how
significant is this difference?
To estimate the significance of the observed deviation,
we utilize a Monte Carlo computer routine.  This
code simulates 22 cluster temperatures between 2.0 and 10.0 keV and
generates velocity dispersions using the virial relation and a $\beta$
value of 1.  It then includes a velocity term for the intrinsic scatter
in the relationship (which is generated by choosing a velocity
perturbation from a uniform distribution of width 150 km s$^{-1}$) as
well as measurement errors in both velocity and temperature (these are
modelled as Gaussians; the dispersion in velocities is 150 km s$^{-1}$
and in temperature is 0.5 keV).  For 1000 simulations, only 40 of the
random datasets had measured
bisector slopes greater than 0.61, the lowest value obtained
for the limited cluster dataset.  The average value for the 1000 runs
was 0.55 $\pm 0.03$.  The highest value of the slope obtained for any of the
simulated datasets is 0.64, which is comparable to the value
obtained for the ES91 dataset but still strongly inconsistent with the
LB93 regression and the limited cluster dataset (with the high
temperatures for A1736 and A3558).

These simulations suggest that while the deviation between the observed
correlation between velocity dispersion and temperature and that predicted
by the virial theorem is small, it is significant.  Clearly
larger individual cluster datasets, higher-quality X-ray spectra, and a
larger dataset of clusters will be vital to improving our understanding of
this fundamental correlation.

The deviation of the $\sigma _r - T$ relationship from that predicted by
the equilibrium model described in Section 1 implies that $\beta$ is a
function of the depth of the gravitational potential, as estimated by
either the temperature or the velocity dispersion.
In this case, defining an average (unweighted) value of
$\beta _{spec}$ for a cluster sample which covers a wide range of
physical parameters yields a quantity which is poorly defined.  The
dependence of $\beta$ on temperature and/or velocity dispersion
is no doubt partially responsible for
the high scatter about the $\sigma _r - T$ relation, which remains even
after elimination of the effects of substructure from the optical dataset.

We have seen in Section 2 that consideration of morphology and substructure
significantly reduces the scatter in the values of $\beta _{spec}$ for the
individual clusters.  Examination of Table 3 reveals that the same effect
does not hold true for the determination of the $\sigma _r - T$ correlation.
Inclusion of the substructure correction actually raises the scatter in the
parameters of the fit slightly,
although it remains comparable to the values obtained
by both ES91 and LB93.  It is clear that
although substructure influences the scatter in the relationship, other
physical effects must also be significant (see also Gerbal et al.\ 1994).

Previous authors have claimed that their data was consistent with the
canonical virial theorem dependence of velocity dispersion on temperature,
$\sigma _r \propto T^{0.5}$ (ES91, LB93).  We have seen that this
``consistency'' is due to the inaccurate use of the least squares
linear regression, and that none of the three datasets are
consistent with the canonical prediction.  Correction for substructure
has very little effect on the slope of the $\sigma _r - T$ correlation.
The scatter to high velocity dispersions implied by the ``steeper
than virial'' relation has been noted
by all previous studies and
generally attributed to velocity substructure. However, we demonstrate
that correction
for substructure has little effect on the correlation.

\section{Mechanisms for Explaining the Discrepancy}

The virial theorem prediction of the relationship between
galaxy velocity dispersion and gas temperature is based on three
assumptions:  that the galaxy orbits are isotropic, that the gas and the
galaxies occupy the same potential well, and that gravity is the only
source of energy for either the gas or the galaxies.  Any process which
may contribute to the deviation of the slope from the virial value must
operate to a different degree in hot, high-$\sigma _v$ clusters than in cooler,
low-$\sigma _v$ systems, to skew the relationship in the observed fashion
(although the effect need not be large).  Mechanisms which have been
proposed include anisotropy in the distribution of galaxy orbits,
incomplete thermalization of the gas, pressure support of the ICM from
magnetic fields, biasing and protogalactic winds.

\subsection{Anisotropy and Magnetic Pressure Support}

The anisotropy parameter $A$ is not well-determined for more than one or
two clusters.
Recall that $A = 1 - \sigma _r ^2 / \sigma ^2 _t$.  For radial orbits,
with $\sigma _r > \sigma _t$, $A < 0$ and $\beta$ is increased (relative to
the value determined by profile fitting; see eqn. 3).  For circularized orbits,
$\sigma _r < \sigma _t$, $A > 0$ and $\beta$ is decreased.  To reproduce the
observed trend in the $\sigma _r - T$ relation, we estimate that hot clusters
require $A \leq -0.1$ (slightly radial orbits), and cool clusters require
$A \sim 0.6$ (moderately circular orbits).  Such an extreme variation in
galaxy anisotropy is not predicted by any current theory of cluster
formation.  Kauffmann \& White (1993) do find some evidence for a
dependence of formation history on mass, but this variation is negligible
over the range of masses included in the limited cluster sample
($5 \times 10^{13} - 1 \times 10^{15}$ M$_{\odot}$; S.\ White, 1994,
private communication).

In most observations, the temperature profile of the ICM is flat out
to the radius where the background dominates the cluster spectrum
(Mushotzky 1994).  Nonetheless, simulations by Evrard (1990)
suggest that the cluster gas will not be completely thermalized after only
one crossing time.  This effect is evident in more detailed calculations by
Metzler \& Evrard (1995, in preparation), who find that the degree of
thermalization is {\it not} systematically dependent on temperature.
Incomplete thermalization clearly affects the distribution of
temperatures measured for the limited cluster sample, but does not
affect the slope of the $\sigma _r-T$
relationship in the required direction.

In an attempt to
resolve the discrepancy between cluster masses determined by gravitational
lensing and those determined from X-rays (Miralda-Escud\'{e} \& Babul 1994),
Loeb \& Mao (1994) propose magnetic pressure support of the intracluster
medium, at least in the cores of cooling flows.  To be dynamically
significant, tangled magnetic fields must contribute a similar amount
of potential energy to the ICM as the gravitational potential.  The required
field strength (on the order of 50 $\mu$G) is large,
but Loeb \& Mao argue that such fields may be generated within
cooling flows, where gas and magnetic field lines are confined and
compressed.

Comparison of the limited cluster sample with Table 1 of Edge, Stewart
\& Fabian 1992 reveals that the majority of the
limited cluster sample possesses cooling flows (as determined from
deprojection analysis)
and therefore may benefit from magnetic pressure support.
Remember, however, that the Loeb \& Mao (1994) analysis
is restricted to the inner 120$h^{-1}$ kpc of A2218 (inside the radius of the
cooling flow), whereas our temperatures
and velocity dispersions are determined for the entire cluster (again
assuming that the cluster ICM temperature profiles are flat outside the
cooling radius, as ASCA data suggest).
It is unclear whether the variation in $\beta$
deriving from magnetic pressure support would be detected in our analysis of
the X-ray and optical data.

\subsection{Protogalactic Winds}

Protogalactic winds provide an additional source of heating of the ICM.
Yahil \& Ostriker (1973), Larson \& Dinerstein (1975) and
White (1991) discuss ram pressure stripping and protogalactic winds
as mechanisms for the metal enrichment of the ICM.  In the winds scenario, the
specific energy of the ICM is affected by the initial collapse of the
cluster, the relative motions of galaxies in the cluster, and winds from
supernova explosions during the formation of elliptical galaxies at early
times.  Of these three physical processes, White (1991) demonstrates that
only protogalactic winds can boost the energy of the gas above the value
determined through the virial theorem.  In addition he shows that the energy
contribution due to winds will be larger in cool clusters than in hot ones.

Using White's Equation 2, we generated a distribution of temperatures
for velocity dispersions ranging from 350-1200 km sec$^{-1}$ (taking his
values for the fraction of intracluster gas coming from winds ($w=0.5$) and
the typical wind velocity in terms of the galactic velocity dispersion
($f_w =3$)).  Fitting these simulated
data, we find that the protogalactic winds model predicts a correlation
between the velocity dispersion and the temperature of a cluster:
\begin{equation}
\sigma _r \propto T^{0.68}
\end{equation}
This depends slightly on the choice of $w$ and $f_w$; for $f_w=2$ we
find that $\sigma _r \propto T^{0.62}$.  The protogalactic wind model
reproduces nearly exactly the dependence of velocity dispersion on
ICM temperature that we find in the limited cluster sample (and which
is consistent with the slopes found by earlier studies).

\subsection{Winds and Biasing}

Another effect which may produce the steepness of the $\sigma _r-T$
relationship is a velocity bias between
cluster galaxies and the background dark matter, which is driven
by dynamical friction (Carlberg 1994; Carlberg \& Dubinski 1991).
Simple virial analysis
predicts $\sigma _r \propto T^{0.5}$ if the collisionless component
has experienced no cooling or heating.
If $\sigma_{DM}$ and $\sigma_{gal}$ refer to the background dark matter
and galaxy velocity dispersions respectively, and assuming the virial
equilibrium holds for the dark matter, then we can write
\begin{equation}
\sigma_{gal} = \sigma_{DM} {\sigma_{gal} \over \sigma_{DM}}
\propto {\sigma_{gal} \over \sigma_{DM}} T^{0.5}
\end{equation}
If the ratio of velocity dispersions is temperature--dependent, then
this will modify the observed $\sigma--T$ relation.

For the purposes of illustration,
we take the distribution of background dark matter velocities to be
Maxwellian,
\begin{equation}
f\left(v\right)\,=\,
{n\left(r\right) \over \left(2\pi\sigma^2\right)^{3/2}}
{\rm exp}\left(-v^2/2\sigma^2\right),
\end{equation}
in which case the Chandrasekhar dynamical friction formula for a galaxy
of mass M in a dark matter potential well with density $\rho$ can be
written as
\begin{equation}
{d{\bf v}_M \over dt}\,=\,
-{4\pi \ln\Lambda G^2 \rho M \over v_M^3}
\left[
{\rm erf}\left(X\right) -
{2X \over \sqrt{\pi} }{\rm exp}\left(-X^2\right)\right] {\bf v}_M,
\end{equation}
with $X\,=\,v_M/\sqrt{2}\sigma$ (Binney and Tremaine 1987).  This can
be rearranged for a characteristic timescale, and writing the bias
for the individual galaxy of mass M, $b\,=\,v_M/\sigma$, we have
\begin{equation}
t_{fric}\,=\,
{ b^3\sigma^3 \over 4\pi \ln\Lambda G^2 \rho M }
\left[{\rm erf}\left(b/\sqrt{2}\right) -
b \sqrt{2 \over \pi} {\rm exp}\left(-b^2/2\right)\right]^{-1}.
\end{equation}
Again, for the purposes of illustration, we assume a power law density
profile for the background dark matter, $\rho\,=\,Ar^{-\alpha}$; then
$\sigma^2 \simeq {GM\left(<R\right) \over R}$ implies
$G\rho \simeq { \left(3-\alpha\right) \over 4\pi R^2} \sigma^2$.
Substituting in, we find that the dynamical friction timescale
for galaxies at a radius R roughly scales as
$t_{fric} \propto \sigma\left(R\right) R^2$.  At a fixed radius R,
more massive (thus typically higher temperature) clusters will have
a higher velocity dispersion, and thus a longer characteristic
timescale for dynamical friction to be significant.  This translates
into a temperature--dependent velocity bias.

Simulations provide an ideal mechanism to test these ideas.  Metzler \&
Evrard (1995) have conducted an ensemble of N--body + hydrodynamic
simulations of the formation and evolution of individual clusters,
explicitly including galaxies and galactic winds.  These simulated
clusters are compared to a ensemble drawn from the same
initial conditions --- but without galaxies and winds --- to isolate
the effects of winds on clusters.  The method is explained in
Metzler \& Evrard (1994).

Figure 2 shows velocity dispersion -- temperature data drawn from
their models.  A ``virial radius'' is identified for each simulated
cluster as the radius with a mean interior overdensity of 170.
The temperatures used are mass--averaged over all gas within the
virial radius; the velocity dispersions are averages drawn from the full
3D velocity information for all dark matter or galaxies within
$r_{vir}$.  A solid line corresponding to $\beta_{spec}\,=\,1$
has also been placed on the plots.

Comparing the dark matter velocity dispersion to the average interior
temperature shows that in the simple two--fluid models, the simulated
clusters are well--fit by the virial relation $\sigma \propto T^{0.5}$.
This is sensible; there is no physics in these models beyond that used
to derive the expected relation.  Note that the values of $\beta_{spec}$
are consistently larger than one; this is a signature of the incomplete
gas thermalization previously seen in other studies.  It is not clear
whether this is physical or numerical in origin; a series of runs with
different resolution would clarify this.

The models including galaxies and winds show different behavior.  Here,
the inclusion of energetic winds, plus dynamical friction of the
galaxy component, provide the necessary physics to deviate from
the virial $\sigma-T$ relation.  For the dark matter, the temperature
dependence is steeper than 0.5, a result of the inclusion of energetic
winds.  When galaxies are used to calculate the velocity dispersion,
however, the relation steepens to $\sigma \propto T^{0.65}$, comparable
to our observed result.  The simulations thus provide evidence for a
temperature--dependent velocity bias,
$\sigma_{gal}/\sigma_{DM} \propto T^{0.1}$.  Both this bias and the
increase in gas temperatures due to energetic winds are responsible
for the final correlation.

It should be noted, of course, that the agreement between the simulated
ensemble and our real clusters is to some degree fortuitous.  The wind model
used in the simulations of Metzler \& Evrard is intentionally of much
greater wind luminosity than expected for real early--type galaxies,
and the dynamical accuracy of modelling galaxies by heavy collisionless
particles in the cluster potential is unclear (Frenk et al.\ 1995).
Nonetheless, this corroborates the theoretical expectation that both
energetic winds and velocity bias can result in the observed
$\sigma-T$ relation.

\section{Discussion}

Although Lubin \& Bahcall (1993) found that the correlation between cluster
velocity dispersion and temperature was somewhat steeper than that predicted
by the virial theorem, the scatter in their dataset was too broad for them
to rule out consistency with the hydrostatic isothermal model.
We show that for our limited dataset,
$\sigma _r \propto T^{0.61 \pm 0.13}$ (GINGA:
$\sigma _r \propto T^{0.76}$), slightly but
significantly (at 96\% confidence) steeper than that
predicted by the virial theorem.  For the ES91 and LB93 datasets, this
discrepancy is significant at the $>99$\% level.  It seems improbable that
this is an artifact of the substructure correction
algorithm.  The mixture modelling technique used to remove substructure
from the cluster datasets does not preferentially
raise the velocity dispersion of high-$\sigma _r$ clusters and
lower that in low-$\sigma _r$ systems, as examination of Table 1
reveals.

The protogalactic winds model of White (1991), in addition to possible velocity
bias due to dynamical friction acting on the cluster galaxies,
quantitatively reproduces
the observed variation in the $\sigma _r - T$ relationship.
Preliminary measurements of cluster emission line diagnostics from
ASCA show metal abundances typical of Type II supernovae, also
supporting the protogalactic winds model (Mushotzky 1994).
(Contrary to the model, however, there is as
yet no conclusive evidence that low-temperature clusters have higher global
abundances than hot systems.)  It seems
plausible that other physical mechanisms, such as velocity anisotropy,
incomplete thermalization of the gas and/or the galaxies, and magnetic
pressure support in cluster cores (which are all likely to be present in
some unknown and variable
degree in clusters) are responsible for the large scatter
about the best-fit $\sigma _r - T$ line.  This scatter is apparent even after
morphology and substructure are considered in the determination of cluster
parameters.

Finally we can relate our revised determination of $\beta _{spec}$ to the
long-standing $\beta$-discrepancy.  Early studies of cluster X-ray
spectroscopy and imaging revealed an important inconsistency:  $\langle
\beta _{spec} \rangle
= 1.2$ (Mushotzky 1984) but $\langle \beta _{fit} \rangle = 0.7$ (Jones \&
Forman 1984).  We have seen that the corrections for morphology and
substructure bring $\langle \beta _{spec} \rangle$ down to 0.9,
only marginally consistent
with $\langle \beta _{fit} \rangle$ (but confirming the earlier results of
ES91).  For many individual clusters,
$\langle \beta _{spec} \rangle$ and $\langle \beta _{fit} \rangle$
are completely different.  Perseus (A426) is the most obvious example,
with $\beta _{spec} = 1.53$ and $\beta _{fit} = 0.57$.
So what is the current status of the
$\beta$-discrepancy?

First of all, we can compare current data on the distribution of gas
and galaxies in clusters.  Schombert (1988) summarizes the data on cluster
density profiles determined from a variety of tracer particles:
\begin{eqnarray}
\rho _{gal} & \propto & r^{-2.6 \pm 0.3} \nonumber \\
\rho _{gas} & \propto & r^{-2.1 \pm 0.2}
\end{eqnarray}
In the hydrostatic isothermal model,
\begin{eqnarray}
\rho _{gas} & \propto & \rho _{gal} ^{\beta _{fit}} \nonumber \\
\beta _{fit} & = & \beta _{spec}
\end{eqnarray}
For our value of $\beta _{spec}$, $\rho _{gas} \propto r^{-2.3}$, which
is at best only marginally consistent with the dependence $\rho _{gas} \propto
r^{-2.1}$ determined by Jones \& Forman (1984).

As Gerbal et al.\ (1994) point out in their theoretical analysis of the
$\beta$-discrepancy, however, in order to test the consistency of the gas
and galaxy scale lengths one must simultaneously observe their radial
dependence {\it independently}, not fitting them together as Jones \&
Forman did.  In the next stage of this project (Bird \& Mushotzky 1995),
we present non-parametric determinations of the galaxy and gas density
profiles based on the MAPEL package (Merritt
\& Tremblay 1994).  MAPEL, a constrained
maximum likelihood algorithm, allows us to determine the best-fit model to
the surface density profiles without assuming a King-model (or other
isothermal) fit to the data (Merritt \& Tremblay 1994).  This is important
because there is growing evidence from gravitational lensing experiments and
computer simulations that the King model fit is not a good description of
the gravitational potential of a galaxy cluster (Navarro, Frenk \& White
1994; see also Beers \& Tonry 1986).
These profiles will
allow us to test on a cluster-by-cluster basis whether the galaxy and gas
profiles differ -- a comparison which in the past has only been possible
in a statistical sense (cf.\ Bahcall \& Lubin 1994).

Note also that in the
time since White (1991) appeared, {\it ROSAT} PSPC and {\it ASCA} surface
density profiles of cool clusters have become publicly available.  These
clusters will be included in the continuation of this project (velocity data
are published in Beers et al.\ 1994).  The protogalactic winds model
predicts that cool clusters will have a larger scale length of gas density
than hot clusters (again, because the relative energy contribution of
winds to the ICM is greater in cool systems).  Use of the expanded dataset
for these clusters will allow us to directly test this prediction and to
probe the effects of protogalactic winds on $\beta _{fit}$.

\acknowledgements

We would like to thank Lori Lubin, Neta Bahcall,
Ray White III, Bill Forman, Christine Jones and
the other attendees of the Aspen Summer Workshop for their contributions to
this
project.  Claude Canizares, Keith Ashman and Alistair Edge also provided useful
conversations during the course of this work.  Andy Fabian's critical
reading of the manuscript greatly improved our statistical analysis.
We are grateful to Simon White
for clarification of issues relating to cluster evolution and
parametrization of cluster density profiles.  This research was supported
in part by NSF EPSCoR grant No.\ OSR-9255223 to the University of Kansas.

\newpage

\begin{table}
\caption{The Cluster Sample}
\begin{tabular}{lcccc} \tableline \tableline
Cluster&$S_{BI}$(uncorr) km s$^{-1}$&$S_{BI}$(corr) km s$^{-1}$&$T_X$ (keV)&
Source Code \\
\tableline
A85&810$^{+76}_{-80}$&810$^{+76}_{-80}$&6.6$^{+1.8}_{-1.4}$&E91 \\
A119&862$^{+165}_{-140}$&1036$^{+214}_{-221}$&5.1$^{+1.0}_{-0.8}$&E91 \\
A193&726$^{+130}_{-108}$&515$^{+176}_{-153}$&4.2$^{+1.6}_{-0.9}$&E91 \\
A194&530$^{+149}_{-107}$&470$^{+98}_{-78}$&2.0$^{+1.0}_{-1.0}$&JF84 \\
A399&1183$^{+126}_{-108}$&1224$^{+131}_{-116}$&6.0$^{+2.1}_{-1.5}$&E91 \\
A401&1141$^{+132}_{-101}$&785$^{+111}_{-81}$&8.6$^{+1.4}_{-1.6}$&E91 \\
A426&1262$^{+171}_{-132}$&1262$^{+171}_{-132}$&6.3$^{+0.2}_{-0.2}$&D93 \\
A496&741$^{+96}_{-83}$&533$^{+86}_{-76}$&4.0$^{+0.06}_{-0.06}$&W94 \\
A754&719$^{+143}_{-110}$&1079$^{+234}_{-243}$&8.7$^{+1.8}_{-1.6}$&E91 \\
A1060&630$^{+66}_{-56}$&710$^{+78}_{-78}$&3.3$^{+0.2}_{-0.2}$&Ikebe 1994 ASCA
\\
A1644&919$^{+156}_{-114}$&921$^{168}_{-141}$&4.1$^{+1.4}_{-0.6}$&E91 \\
A1736$\dagger$&955$^{+107}_{-114}$&528$^{+136}_{-87}$&4.6$^{+0.7}_{-0.6}$&D93
\\
& & & 6.2$^{+0.7}_{-0.7}$&DFER \\
A1795&834$^{+142}_{-119}$&912$^{+192}_{-129}$&5.6$^{+0.1}_{-0.1}$&W94 \\
A2052&1404$^{+401}_{-348}$&714$^{+143}_{-148}$&3.4$^{+0.6}_{-0.5}$&E91 \\
A2063&827$^{+148}_{-119}$&706$^{+117}_{-109}$&3.4$^{+0.35}_{-0.35}$&Yamashita
1992 \\
A2107&684$^{+126}_{-104}$&577$^{+177}_{-127}$&4.2$^{+4.4}_{-1.6}$&D93 \\
A2199&829$^{+124}_{-118}$&829$^{+124}_{-118}$&4.5$^{+0.07}_{-0.07}$&W94 \\
A2634&1077$^{+212}_{-152}$&824$^{+142}_{-133}$&3.4$^{+0.2}_{-0.2}$&D93 \\
A2670&1037$^{+109}_{-81}$&786$^{+203} _{-239}$  &3.9$^{+1.6}_{-0.9}$&D93 \\
A3526&1033$^{+118}_{-79}$&780$^{+100}_{-100}$&3.8$^{+0.3}_{-0.3}$&F94 \\
A3558$\dagger$&923$^{+120}_{-101}$&781$^{+111}_{-98}$&3.8$^{+2.0}_{-2.0}$&D93
\\
& & & 6.2$^{+0.3}_{-0.3}$&DFER \\
DC1842-63&522$^{+98}_{-82}$&565$^{+138}_{-117}$&1.4$^{+0.5}_{-0.4}$&D93 \\
\tableline
\end{tabular}
\end{table}

\newpage

\begin{table}
\caption{$\beta _{spec}$ with and without Substructure Correction}
\begin{tabular}{lccc} \tableline \tableline
Cluster&$\beta _{spec}$(uncorr)&$\beta _{spec}$(corr)& \\
\tableline
A85&0.60&0.60& \\
A119&0.88&1.27& \\
A193&0.76&0.38& \\
A194&0.85&0.67& \\
A399&1.41&1.51& \\
A401&0.92&0.43& \\
A426&1.53&1.53& \\
A496&0.83&0.43& \\
A754&0.36&0.81& \\
A1060&0.73&0.92& \\
A1644&1.25&1.25& \\
A1736&1.20&0.37& {\it Einstein}\\
&0.89&0.27& GINGA\\
A1795&0.75&0.90& \\
A2052&3.51&0.91& \\
A2063&1.22&0.89& \\
A2107&0.67&0.48& \\
A2199&0.92&0.92& \\
A2634&2.07&1.21& \\
A2670&1.67&0.96& \\
A3526&1.70&0.97& \\
A3558&1.36&0.97& {\it Einstein}\\
&0.83&0.60& GINGA\\
DC1842-63&1.18&1.38& \\
\tableline
\end{tabular}
\end{table}

\newpage
\begin{table}
\caption{Fitting the $\sigma _r-T$ Correlation}
\begin{tabular}{lc} \tableline \tableline
Source & Best Fit \\
\tableline
Edge \& Stewart 1991 & $\sigma _r = 10^{2.60 \pm 0.08} T^{0.46 \pm 0.12}$ \\
 $N_{clus}=23$  (pub) &
 $T = 10^{-3.22 \pm 0.77} \sigma _r^{1.35 \pm 0.27}$ \\
Ordinary least squares (no errors) & $\sigma _r = 10^{2.61 \pm 0.06}
T^{0.45 \pm 0.09}$ \\
Bisector (no errors) & $\sigma _r = 10^{2.46 \pm 0.06} T^{0.68 \pm 0.10}$ \\
Bisector (errors) & $\sigma _r = 10^{2.41 \pm 0.51} T^{0.75 \pm 0.08}$ \\
\tableline
Lubin \& Bahcall 1993 & $\sigma _r = 10^{2.53 \pm 0.06} T^{0.62 \pm 0.09}$
(unweighted) \\
 $N_{clus}=41$ (pub)
& $\sigma _r = 10^{2.52 \pm 0.07} T^{0.60 \pm 0.11}$ (weighted$^{\dagger}$) \\
Ordinary least squares (no errors) & $\sigma _r = 10^{2.54 \pm 0.06}
T^{0.61 \pm 0.09}$ \\
Bisector (no errors) & $\sigma _r = 10^{2.38 \pm 0.05} T^{0.84 \pm 0.08}$ \\
Bisector (errors) & $\sigma _r = 10^{2.36 \pm 0.05} T^{0.87 \pm 0.08}$ \\
\tableline
This paper, no substructure correction & $\sigma _r = 10^{2.48 \pm 0.25}
T^{0.73 \pm 0.38}$ \\
$N_{clus}=22$  (bisector with errors)
& $T = 10^{-2.79 \pm 1.54} \sigma _r ^{1.16 \pm 0.52}$ \\
Ordinary least squares (no errors) & $\sigma _r = 10^{2.75 \pm 0.08}
T^{0.31 \pm 0.13}$ \\
Bisector (no errors) & $\sigma _r = 10^{2.51 \pm 0.07} T^{0.69 \pm 0.12}$ \\
\tableline
This paper, substructure correction$\dagger \dagger$
& $\sigma _r = 10^{2.50 \pm 0.09}
T^{0.61 \pm 0.13}$ \\
$N_{clus}=22$ (bisector with errors)
& $T = 10^{-3.15 \pm 0.60} \sigma _r ^{1.31 \pm 0.21}$ \\
Ordinary least squares (no errors)  & $\sigma _r = 10^{2.62 \pm 0.07}
T^{0.42 \pm 0.11}$ \\
Bisector (no errors) & $\sigma _r = 10^{2.45 \pm 0.09} T^{0.69 \pm 0.13}$ \\
\tableline
This paper, substructure correction$\dagger \dagger$
&$\sigma _r = 10^{2.39 \pm 0.09}
T^{0.76 \pm 0.11}$ \\
$N_{clus}=22$ (bisector with errors)
& $T = 10^{-3.21 \pm 0.61} \sigma _r ^{1.32 \pm 0.21}$ \\
\tableline
\end{tabular}
\end{table}

\newpage

\begin{figure}[t]
\vskip -2.0truein
\epsfxsize = 6.0truein
\hskip 1.5truein
\caption{The $\sigma _r - T$ correlation for the limited cluster sample.
Errors in the velocity dispersions (the vertical axis) are taken from
Bird (1994).  Errors in the temperatures are taken from the literature,
identified in Table 1.  The dashed lines are the predicted correlations
for the isothermal $\beta$-model, with $\beta = 1$ or $\beta = 0.67$ and
$\sigma _r \propto T^{0.5}$.
The solid line is the best fit to the data.}
\end{figure}

\begin{figure}[t]
\vskip -2.0truein
\epsfxsize = 6.0truein
\hskip 1.5truein
\caption{The true $\sigma-T$ correlation for the simulated clusters of Metzler
\& Evrard (1995).   Velocity dispersions are the average for all galaxies
or dark--matter particles within an overdensity of 170.  Temperatures
are the mass--average temperature for all gas within an overdensity of
170.  The upper panel shows the results for the ensemble of two--fluid
simulations (without galaxies or energetic winds); here
$\sigma_{DM} \propto T^{0.50}$.  The lower panel shows the results
from the ensemble including galaxies and winds; the crosses show the
$\sigma-T$ relation for the dark matter in these runs, while the boxes
use cluster galaxies.  The results, $\sigma_{DM} \propto T^{0.55}$
and $\sigma_{gal} \propto T^{0.65}$, are steeper than the simple
virial relation.}
\end{figure}

%
%
%
\newpage

\begin{center}
{\bf NOTES TO TABLES}
\end{center}

Table 1.  Source code:  E91 = Edge 1991, JF84 = Jones \& Forman 1984,
DFER = Day et al.\ 1991,
D93 = David et al.\ 1993, W94 = White et al.\ 1994, F94 = Fukuzawa et al.\
1994; $\dagger$ Two-temperature spectral models based on GINGA observations
(Day et al.\ 1991) suggest that these clusters may have higher temperatures
than the {\it Einstein} data suggest.  We have performed our statistical
analysis for both sets of temperatures.

Table 2.  $\dagger$:
LB93 did not published a regression for temperature on velocity
dispersion.  The first regression of velocity dispersion on temperature does
not include weighting by the measurement errors; the second regression is
weighted following a $\chi ^2$ algorithm. $\dagger \dagger$:  The first
set of regressions uses the lower temperatures for A1736 and A3558; the
second set uses the higher temperatures.

\newpage

\begin{references}
\reference{Ashman, K.M. \& Carr, B.J.  1988, \mnras, 234, 219}
\reference{Ashman, K.M., Bird, C.M. \& Zepf, S.E.  1994, \aj, 108, 2348}
\reference{Bahcall, N.A. \& Lubin, L.M.  1994, to appear \apj}
\reference{Beers, T.C., Forman, W., Huchra, J.P., Jones, C. \& Gebhardt, K.
1991, \aj, 102, 1581}
\reference{Beers, T.C., Kriessler, J., Bird, C.M. \& Huchra, J.P.  1995, \aj,
109, to appear March}
\reference{Beers, T.C. \& Tonry, J.L.  1986, ApJ, 300, 557}
\reference{Binney, J. \& Tremaine, S. 1987, {\it Galactic Dynamics} (Princeton:
Princeton University Press)}
\reference{Bird, C.M.  1993, Ph.D thesis, University of Minnesota and Michigan
State University}
\reference{Bird, C.M.  1994, \aj, 107, 1637}
\reference{Bird, C.M. \& Mushotzky, R.F.  1995, in preparation}
\reference{Buote, D.A. \& Canizares, C.R.  1992, \apj, 400, 385}
\reference{Carlberg, R.G.  1994, \apj, 433, 468}
\reference{Carlberg, R.G. \& Dubinski, J.  1991, \apj, 369, 13}
\reference{Davis, D.S. \& Mushotzky, R. F. 1993, \aj, 105, 491}
\reference{David, L.P., Slyz, A., Jones, C., Forman, W., Vrtilek, S.D. \&
Arnaud, K.  1993, \apj, 412, 479}
\reference{Day, C.S.R., Fabian, A.C., Edge, A.C. \& Raychaudhury, S.  1991,
\mnras, 252, 394}
\reference{Dressler, A.  1984, \araa, 22, 185}
\reference{Edge, A.C.  1991, \mnras, 250, 103}
\reference{Edge, A.C. \& Stewart, G.C.  1991, \mnras, 252, 428}
\reference{Edge, A.C., Stewart, G.C. \& Fabian, A.C.  1992, \mnras,
258, 177}
\reference{Evrard, A.E.  1990, in {\it Clusters of Galaxies}, eds.
Oegerle, W.R., Fitchett, M.J. \& Danly, L., (New York:  Cambridge University
Press), 287}
\reference{Fabricant, D.M., Lecar, M. \& Gorenstein, P.  1980, \apj,
241, 552}
\reference{Frenk, C.S., Evrard, A.E., White, S.D.M., \& Summers, F.J.
1995, \apj, submitted}
\reference{Fukuzawa, Y., Ohashi, T., Fabian, A.C., Canizares, C.R.,
Ikebe, Y., Makishima, K., Mushotzky, R.F. \& Yamashita, K.  1994, \pasj,
46, L55}
\reference{Gerbal, D., Durret, F. \& Lachi\`{e}ze-Rey, M.  1994, \aa,
288, 746}
\reference{Giovanelli, R. \& Haynes, M.  1985, \apj, 292, 404}
\reference{Jones, C. \& Forman, W.  1984, \apj, 276, 38}
\reference{Kauffmann, G. \& White, S.D.M.  1993, \mnras, 261, 921}
\reference{Kent, S.M. \& Gunn, J.E.  1982, 87, 945}
\reference{Kent, S.M. \& Sargent, W.L.W.  1983, 88, 697}
\reference{Larson, R.B. \& Dinerstein, H.L.  1975, \pasp, 87, 911}
\reference{Loeb, A. \& Mao, S.  1994, \apjl, 435, L109}
\reference{Lubin, L.M. \& Bahcall, N.A.  1993, \apjl, 415, L17}
\reference{McLachlan, G.J. \& Basford, K.E.  1988, {\it Mixture Models},
(New York:  Marcel Dekker)}
\reference{Merritt, D.  1987, \apj, 313, 121}
\reference{Merritt, D. \& Tremblay, B.  1994, \aj, 108, 514}
\reference{Metzler, C.A. \& Evrard, A.E.  1994, \apj, 437, 564}
\reference{Metzler, C.A. \& Evrard, A.E.  1995, \apj, in preparation}
\reference{Miralda-Escud\'{e}, J. \& Babul, A.  1994, preprint}
\reference{Mohr, J., Fabricant, D. \& Geller, M.  1993, CfA preprint}
\reference{Mushotzky, R.F.  1984, Phys.\ Scripta, T7, L157}
\reference{Mushotzky, R.F.  1994, {\it New Horizons}, (Tokyo:
publisher unknown), in press}
\reference{Navarro, J., Frenk, C. \& White, S.D.M.  1994, \mnras, in press}
\reference{Sarazin, C.L.  1988, {\it X-ray emissions from clusters of
galaxies}, (Cambridge:  Cambridge University Press)}
\reference{Schombert, J.M.  1988, \apj, 328, 475}
\reference{Sharples, R.M., Ellis, R.S. \& Gray, P.M.  1988, \mnras,
231, 479}
\reference{White, R.E. III.  1991, \apj, 367, 69}
\reference{White, R.E. III, Day, C.S.R., Hatsukade, I. \& Hughes, J.P.
1994, \apj, 433, 583}
\reference{Yahil, A. \& Ostriker, J.P.  1973, \apj, 185, 787}
\reference{Yamashita, K.  1992, {\it Frontiers of X-ray Astronomy},
(Tokyo:  publisher unknown), 473}
\end{references}
\end{document}